\documentclass[aps,prb,twocolumn,superscriptaddress,showpacs,preprintnumbers,amsmath,amssymb,floatfix]{revtex4}
\usepackage{hyperref}
\usepackage{epsfig}
\usepackage{graphicx}
\usepackage{dcolumn}
\usepackage{amsfonts}
\usepackage{amsmath}
\usepackage{dsfont}
\usepackage[usenames]{color}

\begin{document}

\title{Optimized single-qubit gates for Josephson phase qubits}
\date{\today}

\author{Shabnam Safaei}
\email{safaei@sns.it}
\affiliation{NEST-CNR-INFM and Scuola Normale Superiore, Piazza dei Cavalieri 7, I-56126 Pisa, Italy}
\author{Simone Montangero}
\affiliation{Institut f\"ur Quanteninformationsverarbeitung, Universit\"at Ulm, D-89069 Ulm, Germany}
\affiliation{NEST-CNR-INFM and Scuola Normale Superiore, Piazza dei Cavalieri 7, I-56126 Pisa, Italy}
\author{Fabio Taddei}
\author{Rosario Fazio}
\affiliation{NEST-CNR-INFM and Scuola Normale Superiore, Piazza dei Cavalieri 7, I-56126 Pisa, Italy}

\begin{abstract}
In a Josephson phase qubit the coherent manipulations of the computational states are 
achieved by modulating an applied ac current, typically in the microwave range. In this work 
we show that it is possible to find optimal modulations of the bias current to achieve 
high-fidelity gates. We apply quantum optimal control theory to determine the form of the pulses 
and study in details the case of a NOT-gate. To test the efficiency of the optimized pulses in 
an experimental setup, we also address the effect of possible imperfections in the pulses shapes, 
the role of off-resonance elements in the Hamiltonian, and the effect of capacitive interaction 
with a second qubit. 
\end{abstract}

\pacs{85.25.Cp, 03.67.Lx, 02.30.Yy}
\maketitle

\section{Introduction}
\label{intro}

Over the past decades, together with the development of the theory of quantum information~\cite{Nielsen} 
there has been an increasing effort to find those physical systems where quantum information 
processing could be implemented.  Among the many different proposals, devices based on superconducting 
Josephson junctions are promising candidates in the solid state realm (see the reviews in Ref.
~\onlinecite{Makhlin01,Wendin07,you05,clarke08}).  Josephson qubits can be categorized into three main 
classes: Charge, phase and flux qubits, depending on which dynamical variable is most well defined 
and consequently which basis states are used as computational states $|0\rangle$ and $|1\rangle$.

Phase qubits~\cite{Martinis02,Martinis03,Claudon04}, subject of the present investigation,
in their simplest configuration can be realized with a single current biased Josephson 
junction. For bias lower that the critical current, the two lowest eigenstates of the system 
form the computational space. The application of a current pulse, with frequency 
which is in resonance with the transition frequency of the two logical states, typically in the 
microwave range, allows to perform all desired single bit operations.
Recent experiments~\cite{SteffenSCI06,SteffenPRL06} have realized both single-bit and two-bit 
gates in capacitive coupled phase qubits. In the experiments conducted so far, motivated by similar 
approach in NMR, the amplitude of the microwave current used to perform the qubit manipulation has
a Gaussian shape~\cite{SteffenPRL06,Steffen03}. The importance of achieving fast quantum gates with 
high fidelity rises the question whether there are modulations, other than Gaussian, 
which lead to higher fidelities. Indeed, Gaussian pulses act better when their 
duration time is longer \cite{Steffen03}, therefore the search for modulations which 
result in high fidelity gates, even when the duration time of the pulse is short, seems to be 
necessary. Some theoretical work has already been done in this direction to examine the efficiency 
of different modulations~\cite{Steffen03}. In the present paper we follow a different approach 
as compared to Ref. \onlinecite{Steffen03} and show that by employing the quantum optimal control 
theory \cite{Peirce88, Borzi02, Sklarz02,Calarco04}, we can further improve the (theoretical) 
bounds on the error of gate operations. 

Quantum optimal control has been already applied to optimize quantum manipulation of 
Josephson nanocircuits in the charge limit~\cite{Sporl07,Montangero07,Safaei08}. 
Here we want to test this method in the opposite regime of phase qubit~\cite{footnote} 
and see whether it is possible to find optimal modulations of microwave pulses, with 
different duration times, which give very good fidelity for single bit operations. 

The paper is organized as follows: 
in Sec.~\ref{model} we will describe the model for the phase qubit and the Hamiltonian 
used in the rest of the paper. In Sec.~\ref{gate} we introduce the NOT quantum gate
which we have chosen to optimize. Then a brief introduction to the quantum optimal control 
algorithm which is used for this work will be given in Sec.~\ref{qoc}. The numerical results 
for a phase qubit  will be presented in  Sec.~\ref{res}. The achieved accuracy for desired 
operation, discussed in Sec.~\ref{optimal}, is further tested against possible imperfections 
in the pulses shape (Sec.~\ref{shapes}), presence of off-resonance elements in the 
Hamiltonian (Sec.~\ref{off}) and possible presence  of the inter-qubit capacitive 
interaction in multi-qubit systems (Sec.~\ref{int}). The specific question of the leakage out of 
the Hilbert space is addressed in Sec.~\ref{five}, where we provide numerical results obtained for a 
junction with five levels inside its potential. A summary of the results obtained and 
possible perspectives of this work  will be presented in the concluding remarks in Sec.~\ref{conc}.

\section{Single-junction phase qubit}
\label{model}

\begin{figure}
\begin{center}
\includegraphics[width=1.0\linewidth]{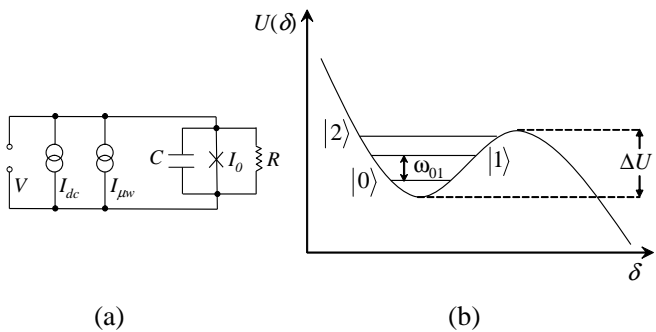}
\caption{(a) Schematic drawing of a single-junction phase qubit with capacitance $C$, 
resistance $R$ and critical current $I_0$ which is biased by a $dc$ current $I_{dc}$. 
A microwave pulse $I_{\mu w}= I(t)\cos(\omega t+\varphi)$, with frequency $\omega=\omega_{01}$, 
is applied to make transitions between the two lowest energy levels of the system $|0\rangle$ and $|1\rangle$. 
(b) By neglecting the resistive branch, the potential energy of the system $U$,
as a function of the Josephson phase across the junction $\delta$, has the form of a tilted washboard 
potential. This potential is defined by the height of the well $\Delta U$ and the frequency of 
the classical oscillations in the bottom of the well.  
\label{fig:junction}}
\end{center}
\end{figure}

A phase qubit can be realized by a flux-biased rf SQUID~\cite{Simmonds04}, a low inductance 
dc SQUID~\cite{Claudon04} or a large inductance dc SQUID~\cite{Martinis02}.
In its simplest design a phase qubit consists of 
a single Josephson junction (Fig.~\ref{fig:junction}(a)) with critical current $I_0$ 
and a biasing dc current $I_{dc}$. 
The Hamiltonian has the form
\begin{eqnarray}
\label{eq:washboard}
H_{dc} = - E_C \frac{{\partial}^2}{\partial{\delta}^2}
- E_J \cos(\delta) - \frac{I_{dc} \Phi_0}{2 \pi}~\delta,
\end{eqnarray}
where $E_C = (2e)^2 / 2C$ and $E_J = I_0 \Phi_0 / 2 \pi$ are, respectively,
the charging energy and the Josephson energy of the junction with capacitance $C$,
$\Phi_0 = h/2e$ being the quantum of flux, and $\delta$ represents the Josephson phase 
across the junction. The regime in which the superconducting phase $\delta$ is the 
appropriate quantum variable is reached when $E_J \gg E_C$. The potential energy of 
the system, as a function of $\delta$, has the form of tilted washboard with 
quantized energy levels inside each well (Fig.~\ref{fig:junction}(b)). 
When $I_{dc}\lesssim I_0$ there are few levels inside each well and the two 
lowest states $|0\rangle$ and $|1\rangle$, with energies $E_0$ and $E_1$ and transition 
frequency $\omega_{01}=(E_1-E_0)/\hbar\approx 5$GHz, can be used as computational states. 
The transition between the two lowest states is made by use of a microwave current 
$I_{\mu w}$ of frequency $\omega$ which is in resonance with transition frequency 
$\omega_{01}$. Transition to higher states ($|2\rangle,~|3\rangle~...$), which are out 
of qubit manifold, are off-resonance due to the anharmonicity of the potential well. 
 
To be effectively used as a two-level quantum system, the junction is biased with a dc current slightly 
smaller than the critical current $I_{dc}\lesssim I_0$. In this regime the potential energy 
of the system can be approximated by a cubic potential and the Hamiltonian 
(\ref{eq:washboard}) becomes:
\begin{eqnarray}
\label{eq:cubic}
H_{dc} &\approx& - E_C \frac{{\partial}^2}{\partial{\delta}^2} - \frac{\Phi_0}{2\pi}
(I_0 - I_{dc})(\delta - \frac{\pi}{2})\nonumber\\ &-& \frac{I_0 \Phi_0}{12\pi} 
(\delta - \frac{\pi}{2})^3.
\end{eqnarray}
The application of a  microwave current $I_{\mu w}=I(t)\cos(\omega t + \varphi)$ is taken into account 
by adding the linear term $H_{\mu w}=\frac{\Phi_0}{2\pi} I_{\mu w} \delta$ to 
the Hamiltonian (\ref{eq:washboard}). Since the eigenstates of the junction biased 
with a dc current are used as computational states, it is appropriate to write the 
full Hamiltonian in the basis of the eigenstates $|n \rangle$ of the 
system with dc bias current.  To examine the effect of microwave current, one needs
to know the elements of the superconducting phase $\delta$ in this basis. 

Moving to the rotating frame, in which the fast 
oscillations due to $\cos(\omega t + \varphi)$ do not appear,
the Hamiltonian $\tilde{H}$ in the rotating frame is related to the 
Hamiltonian in laboratory frame $H$ via
\begin{eqnarray}
\label{eq:lab-to-rot}
\tilde{H}=VHV^{\dag}-i\hbar V\frac{\partial}{\partial t}V^{\dag},
\end{eqnarray}
whereas the state of the system in the rotating frame is $|\tilde{\psi}\rangle=V |\psi\rangle$.
By introducing $g(t)=I(t)\sqrt{\hbar/2 C \omega_{01}}$ and $\Delta_{mn}=
\frac{1}{2} \sqrt{\hbar\omega_{01}} \langle m|\delta|n\rangle$, 
and considering only the first three levels in 
the well, the Hamiltonian of the phase qubit in the rotating frame takes the following form
\begin{eqnarray}
\label{eq:rotating2}
\tilde{H} \approx
\left(\begin{matrix}
0&g(t)\Delta_{01}e^{i\varphi}&0\\
g(t)\Delta_{10}e^{-i\varphi}&0&g(t)\Delta_{12}e^{i\varphi}\\
0&g(t)\Delta_{21}e^{-i\varphi}&-\hbar\delta\omega
\end{matrix}\right) .
\end{eqnarray}
Here we have set $E_0=0,~\omega=\omega_{01}$, $\delta\omega\equiv\omega_{01}-\omega_{12}$ 
and we have assumed that off-resonance terms have negligible effect.
As we shall see, by a proper choice of $\varphi$ and microwave current modulation $g(t)$ 
it is possible to perform single-bit operations on the computational states 
$|0\rangle$ and $|1\rangle$.

\section{NOT-gate}
\label{gate}

As one can see from the $2\times2$ top-left block of the Hamiltonian 
(\ref{eq:rotating2}), the initial phase of the microwave pulse $\varphi$ 
defines the axis of rotation, in $xy$-plane of the Bloch sphere, for a given state, 
while the pulse amplitude and duration time define the angle of rotation.
For example, by setting $\varphi=0$ ($\varphi=\pi/2$) such block is 
proportional to the Pauli matrix $\sigma_x$ ($\sigma_y$), {\em i. e.} a 
rotation around the $x$($y$)-axis.  
In a recent experiment~\cite{SteffenSCI06} a $\pi$ rotation around $x$ has 
been implemented as a part of a sequence of operations to create entanglement between two 
phase qubits.
This motivates us to set $\varphi=0$ and focus this work on the single-qubit 
NOT-gate operation consisting of a $\pi$ rotation around the $x$-axis. 

In the typical experiment a shaped pulse with the following Gaussian modulation 
\cite{Steffen03}
\begin{eqnarray}
\label{eq:gaussian}
g(t)=\frac{a}{t_g}~e^{-\frac{(t-\alpha t_g)^2}{2t_g^2}}
\end{eqnarray}
is used to induce flips between states $|0\rangle$ and $|1\rangle$ and vice versa.
Here $a$, $t_g$ and $T=2\alpha t_g$ are, respectively, the amplitude, 
characteristic width and total width of the pulse, $\alpha$ being the cut-off of 
the pulse in time. The actual result of the operation can be quantified by the 
fidelity $|\langle\psi(T)|\psi_{fin}\rangle|^2$, where $|\psi_{fin}\rangle$ is the 
desired final state and $|\psi(T)\rangle$ is the state achieved at the end of time 
evolution starting from initial state $|\psi(t=0)\rangle=|\psi_{ini}\rangle$.

\begin{figure}[top]
\begin{center}
\includegraphics[width=0.9\linewidth]{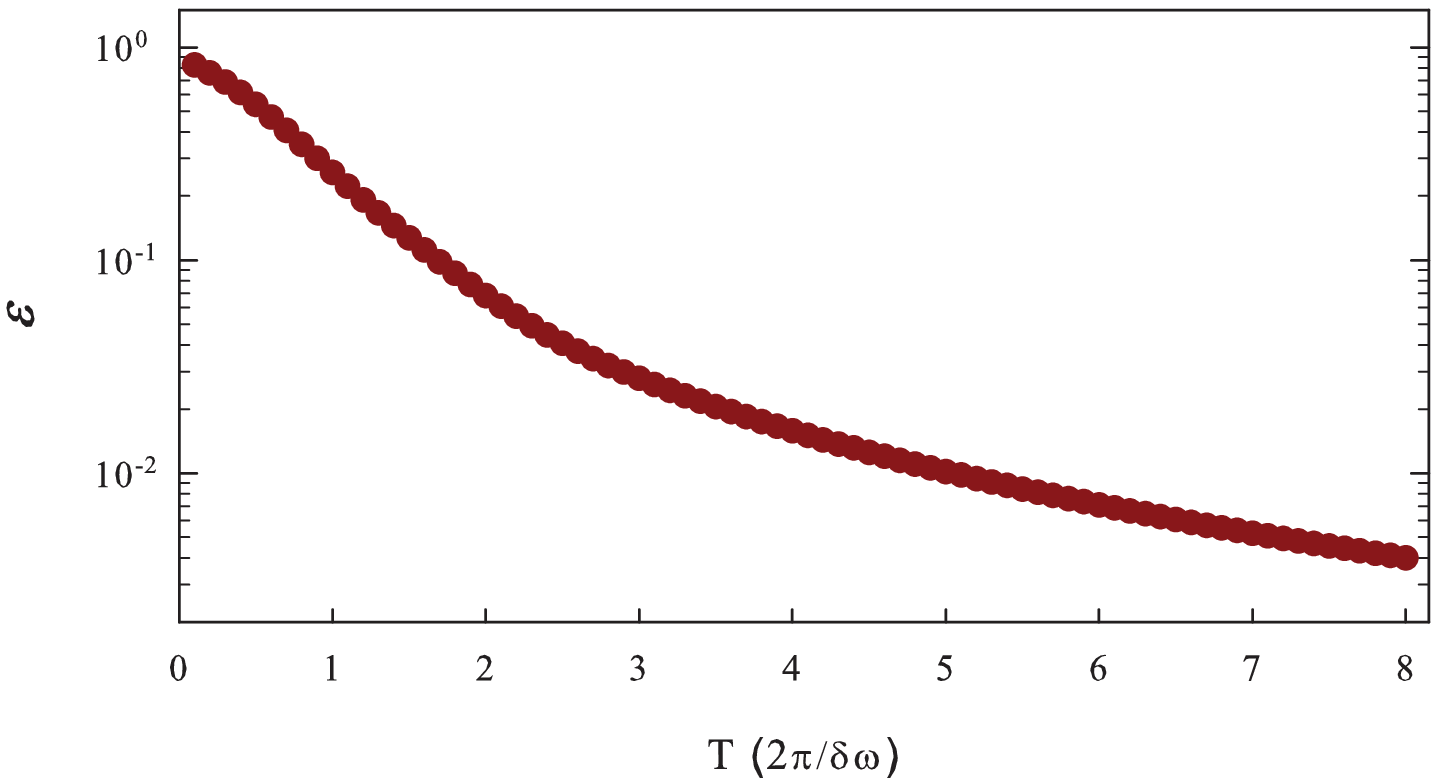}
\caption{(Color on line) Error $\cal E$ for a NOT-gate operation 
on an arbitrary superposition of states $|0\rangle$ and $|1\rangle$ 
after applying a Gaussian pulse with amplitude $a=1.25$, cut-off 
$\alpha=3$ as a function of the duration time $T$.  
\label{fig:error-example}}
\end{center}
\end{figure}

\begin{figure}[top]
\begin{center}
\includegraphics[width=0.9\linewidth]{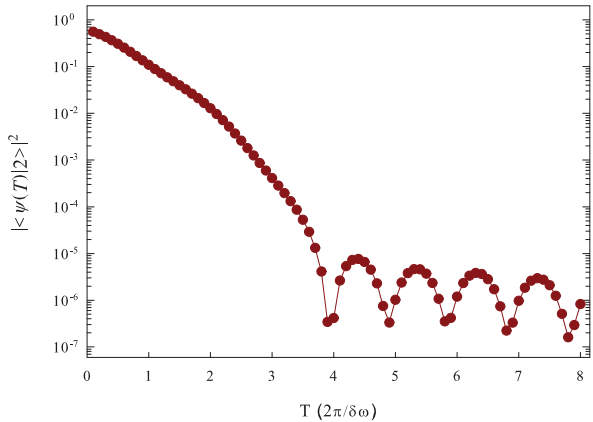}
\caption{(Color on line) The leakage outside the qubit manifold 
for a NOT-gate operation on an arbitrary superposition of states 
$|0\rangle$ and $|1\rangle$ after applying a Gaussian pulse 
with amplitude $a=1.25$, cut-off $\alpha=3$ as a function of the duration time $T$.  
\label{fig:leakage-example}}
\end{center}
\end{figure}

For a $\pi$ rotation and with a typical cut-off value ($3\leq\alpha\leq5$) 
the amplitude $a\approx \sqrt{\pi/2}$ yields a pretty high fidelity of 
rotation. More precisely, Fig.~\ref{fig:error-example} shows the error 
${\cal{E}}=1-|\langle\psi(T)|\psi_{fin}\rangle|^2$ for a NOT-gate 
operation on an arbitrary superposition $|\psi_{ini}\rangle=b|0\rangle+c|1\rangle$, 
which would result in the state $|\psi_{fin}\rangle=b|1\rangle+c|0\rangle$, using a 
Gaussian pulse with cut-off $\alpha=3$, amplitude $a=1.25$ and duration time $T$.
The leakage outside the qubit manifold, defined as $|\langle\psi(T)|2\rangle|^2$, 
is also shown in Fig.~\ref{fig:leakage-example}. It is worthwhile noting that although 
the leakage, for long enough pulses, can be of the order of $10^{-7}$, 
the error in the NOT-gate operation is always higher than $10^{-3}$~
(see Ref.~\onlinecite{fazio99}).

\section{Quantum optimal control}
\label{qoc}

As we mentioned in the Introduction, in this work we use quantum optimal 
control theory in order to obtain microwave current modulations
which give rise to a high-fidelity NOT-gate operation for a phase qubit. 
In this section we briefly review the optimal control algorithm which we 
have employed to obtain optimized modulations.

In general, quantum optimal control algorithms~\cite{Peirce88, Borzi02, 
Sklarz02} are designed to lead a quantum system with state 
$|\psi(t)\rangle$ from an initial state $|\psi(0)\rangle=|\psi_{ini}\rangle$
to a target final state $|\psi_{fin}\rangle$ at time $T$ by minimizing a cost 
functional which is a measure of inaccuracy of reaching the desired final state.
If $|\psi(T)\rangle$ denotes the state achieved at time $T$, one can consider 
two different cost functionals:
\begin{itemize}
\item $e_1=1-|\langle\psi(T)|\psi_{fin}\rangle|^2$

By minimizing this cost functional, although the population of the 
desired state $|\psi_{fin}\rangle$ will be maximized, the overall 
phase of this state is not forced to be preserved.

\item $e_2=\left\||\psi(T)\rangle-|\psi_{fin}\rangle\right\|^2$

Minimization of this second cost functional, in addition to maximizing 
the population of the desired state, preserves its overall phase. 
\end{itemize}

In optimal control theory the minimization of the cost functional is done by 
updating the Hamiltonian of the system, via some control parameters, in an 
iterative procedure until the desired value of the cost functional is reached. 
Any specific algorithm which is guaranteed to give improvement at each 
iteration~\cite{ref:Sola98} is called \textit{immediate feedback control} 
and can be briefly described as follows: 
Assume that the Hamiltonian of the system depends on a set of parameters 
$\{u_{j}(t)\}$ which are controllable. By using a proper initial guess 
$\{u^{(0)}_{j}(t)\}$ for control parameters, first the state of the 
system $|\psi(t)\rangle$ is evolved in time with the initial condition 
$|\psi(0)\rangle=|\psi_{ini}\rangle$ giving rise to $|\psi(T)\rangle$ 
after time $T$. At this point the iterative algorithm starts, aiming at 
decreasing the cost functional by adding a correction to control parameters 
in each step. In the $n$th step of this iterative algorithm 
\begin{itemize}
\item An auxiliary state $|\chi(t)\rangle$ is evolved backward in time 
starting from $|\chi(T)\rangle$ reaching $|\chi(0)\rangle$.

In the case of minimizing $e_1$, 
$|\chi(T)\rangle=|\psi_{fin}\rangle\langle\psi_{fin}|\psi(T)\rangle$
and for minimizing $e_2$, 
$|\chi(T)\rangle=2(|\psi(T)\rangle-|\psi_{fin}\rangle)$.

\item The states $|\chi(0)\rangle$ and $|\psi(0)\rangle$ are evolved forward in time, 
respectively, with control parameters $\{u^{(n)}_{j}(t)\}$ and $\{u^{(n+1)}_{j}(t)\}$. 
Here,
\begin{eqnarray}
\label{eq:update}
u^{(n+1)}_{j}(t)=u^{(n)}_{j}(t)+\frac{2}{\lambda(t)}\Im\left[\langle\chi(t)|\frac{\partial H}
{\partial u_{j}(t)}|\psi(t)\rangle\right]
\end{eqnarray}
are updated control parameters. $\lambda(t)$ is a weight function used to fix initial 
and final conditions on the control parameters in order to avoid major changes at the 
beginning and end of time evolution and is an important parameter for the convergence of 
the algorithm.
\end{itemize}
These two steps are repeated until the desired value of $e_1$ or $e_2$ is obtained.

In order to implement the optimization procedure to a NOT-gate for 
any arbitrary superposition of computational states, one must be able to flip $|0\rangle$ 
and $|1\rangle$ at the same time ({\em i. e.} with same pulse) making sure that the phase 
relation between them is preserved. This is guaranteed by using the following definition of 
fidelity 
\begin{eqnarray}
\label{eq:not-fidelity}
{\cal{F}}\equiv\left|\frac{\langle\psi_0(T)|1\rangle+\langle\psi_1(T)|0\rangle}{2}\right|^2,
\end{eqnarray}
where $|\psi_0(T)\rangle$ and $|\psi_1(T)\rangle$ are final states achieved at time $T$ 
after applying the same pulse on initial states $|0\rangle$ and $|1\rangle$.
The minimization of the cost functional $e_1$, for flipping at the same time the states 
$|0\rangle$ and $|1\rangle$, does not necessarily lead to maximization of the fidelity 
(\ref{eq:not-fidelity}) due to possible changes in the phase relation between them. 
However if $e_2$ is minimized, the maximal fidelity is also guaranteed. Therefore in order 
to obtain a high-fidelity NOT-gate it seems more natural to minimize $e_2$ instead of $e_1$. 
However, in the following we will show that, although in the ideal case optimized pulses 
obtained from minimizing $e_2$ result in much higher fidelity, when more realistic 
cases are considered optimized pulses from minimizing $e_1$ lead to higher 
fidelities, specially for very short pulses. In this work we often use the error 
${\cal{E}}=1-{\cal{F}}$ instead of fidelity.

\section{Numerical Results}
\label{res}

In this section we present the numerical results to show that the 
quantum optimal control theory allows 
to optimize the modulation of microwave pulses in order to implement a 
high-fidelity NOT-gate. The optimization is done in the rotating frame 
and the Hamiltonian (\ref{eq:rotating2}) is used for time evolution while 
$\Delta_{ij}$ are calculated by means of perturbation theory.

\subsection{Optimal NOT-gate}
\label{optimal}

\begin{figure}[top]
\begin{center}
\includegraphics[width=1.0\linewidth]{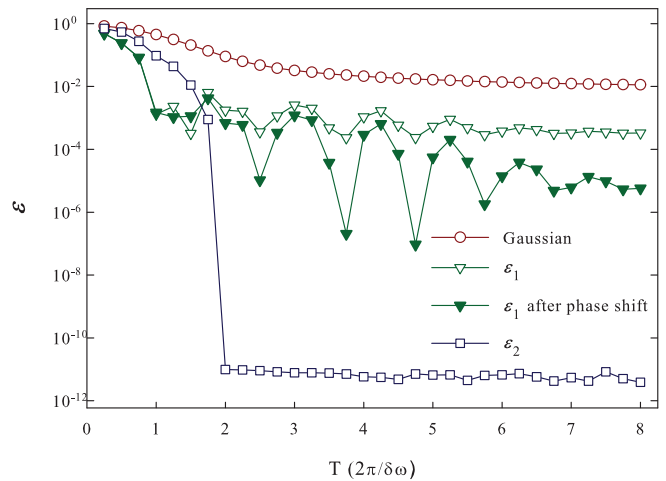}
\caption{(Color on line) Error for a NOT-gate operation applied to any 
arbitrary superposition of states $|0\rangle$ and $|1\rangle$ made by 
pulses with Gaussian modulation (circles) and optimized modulation obtained 
from minimizing $e_1$ (unfilled triangles) and $e_2$ (squares) in a three-level 
system. Gaussian pulses have amplitude $a=1.25$ and cut-off in time $\alpha=3$.
Optimized pulses are obtained after at most 5000 iterations and using Gaussian 
pulses as initial guess. Filled triangles are obtained by applying a $0.01\pi$ 
phase shift after optimized pulses obtained from minimizing $e_1$.
\label{fig:optimized}}
\end{center}
\end{figure}

\begin{figure}[top]
\begin{center}
\begin{tabular}{cc}
\includegraphics[width=1.0\linewidth]{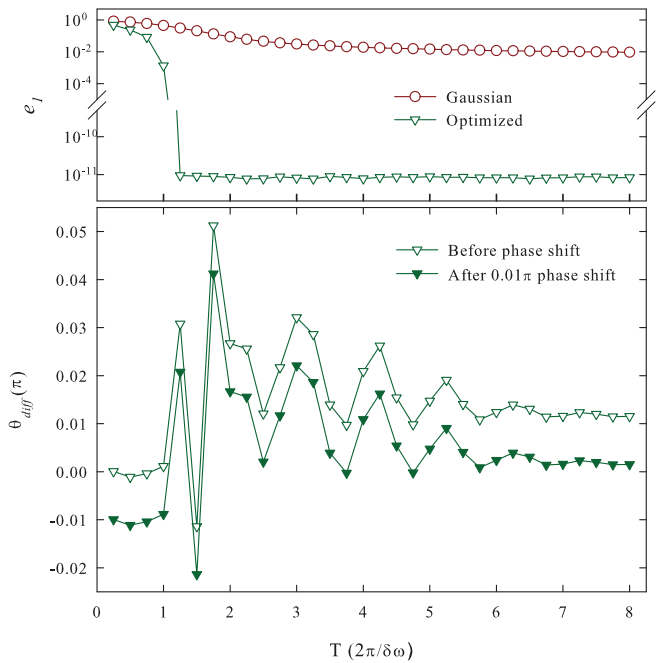}
\end{tabular}
\end{center}
\caption{(Color on line) Top panel: the averaged value of $e_1$ for 
$|0\rangle\leftrightarrow|1\rangle$ transitions after applying pulses 
with Gaussian modulation (circles) and optimized modulation (triangles). 
Optimized pulses  are obtained by minimizing $e_1$. Vertical axis is in 
logarithmic scale and $T$ is the total width of the pulse. 
Bottom panel: the phase difference between final $|0\rangle$ and $|1\rangle$ 
states (after applying the optimized pulses) in units of $\pi$. This final 
phase difference increases the error of NOT-gate ${\cal{E}}_1$ to what has 
been shown in Fig.~\ref{fig:optimized}. In principle, a proper phase shift 
gate can compensate this phase difference and decrease the error to $10^{-12}$.  
\label{fig:phase-diff}}
\end{figure}

By employing the quantum optimal control algorithm described in the 
Sec.~\ref{qoc} and using the modulation of the microwave pulse $g(t)$ 
as the control parameter, we start from Gaussian pulses (\ref{eq:gaussian})
of given duration time $T$ as the initial guess and optimize the 
NOT-gate operation. We will show the results obtained from minimizing both 
$e_1$ and $e_2$ and refer to corresponding errors by ${\cal{E}}_1$ and ${\cal{E}}_2$ 
and corresponding optimized pulses by $g_1$ and $g_2$. The optimization has 
been stopped when either the cost functionals reached the value $10^{-12}$ 
or 5000 iterations are done.

Figure~\ref{fig:optimized} shows the error $\cal{E}$, as a 
function of duration time of the pulse $T$, for the Gaussian pulses 
used as initial guess (circles) and for the optimized pulses 
(unfilled triangles and squares). For most of points, the 
convergence is reached in much less than 5000 iterations. However 
for pulses with $T<2~\frac{2\pi}{\delta\omega}$, 5000 iterations has been completed.
As we expected, minimizing $e_2$ results in high-fidelity NOT-gate with 
${\cal{E}}\approx 10^{-12}$ for all $T\geq 2~\frac{2\pi}{\delta\omega}\approx 4$ ns, while 
for very short pulses it seems that, with same number of iterations, minimizing 
$e_1$ leads to better results.  

In order to understand the reason for the oscillating behavior of ${\cal{E}}_1$ as a 
function of $T$, we plot the average value of $e_1$ for $|0\rangle\rightarrow|1\rangle$ 
and $|1\rangle\rightarrow|0\rangle$ transitions at the end of optimization (top panel 
of Fig.~\ref{fig:phase-diff}) which shows that the final value of $e_1$ for both of 
these transitions is of the order of $10^{-12}$. As we explained in Sec. 
\ref{qoc}, 
$e_1$ is insensitive to the phase of the final state and it turns out that while a 
given optimized pulse applied to initial state $|0\rangle$ leads to the final state 
$e^{i\theta_0}|1\rangle$, then the same pulse might transform the initial state $|1\rangle$ 
into $e^{i\theta_1}|0\rangle$, {\em i. e.} there is a phase difference between the 
two final states $\theta_{diff}\equiv\theta_1-\theta_0$. The bottom panel of 
Fig.~\ref{fig:phase-diff} shows this phase difference for optimized pulses with given 
duration time $T$ which increases the error of NOT-gate ${\cal{E}}_1$ to what has been 
shown in Fig.~\ref{fig:optimized}. 

Although this phase difference causes a major increase in the error while working 
with superpositions, the error ${\cal{E}}_1$ is at least one order of magnitude 
smaller than those from Gaussian pulses (Fig.~\ref{fig:optimized}). 
Moreover the final phase difference between $|0\rangle$ and $|1\rangle$ can be 
compensated by a following phase shift gate. In Fig.~\ref{fig:optimized} results after 
applying a $0.01\pi$ (which is approximately the average of $\theta_{diff}$ in time) 
phase shift are also shown (filled triangles) which demonstrate a significant 
decrease of ${\cal{E}}_1$.

\subsection{Imperfections in the pulse shapes}
\label{shapes}

\begin{figure}[top]
\begin{center}
\begin{tabular}{cc}
\includegraphics[width=0.9\linewidth]{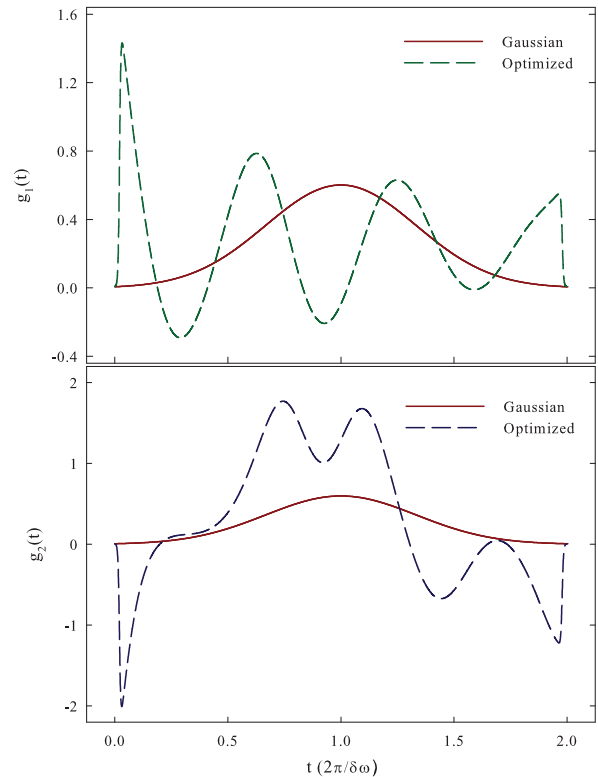}
\end{tabular}
\end{center}
\caption{(Color on line) Examples of final optimized modulation of pulses 
(dashed lines) and the corresponding pulse with Gaussian modulation (solid lines) 
used as initial guess in optimization process with duration time 
$T=2~\frac{2\pi}{\delta\omega}$ obtained from minimizing $e_1$ (top panel) and $e_2$ 
(bottom panel).
\label{fig:pulses}}
\end{figure}

In this section we study the Fourier transform of the optimized pulses,
in order to see how practically they are realizable in the laboratory,
and to examine the effect of high-frequency components. Two examples of 
the final optimized pulses (dashed lines) are shown in Fig.~\ref{fig:pulses}, 
both with duration time $T=2~\frac{2\pi}{\delta\omega}\approx 4$ ns. 
Optimized $g_1$ (top panel) and $g_2$ (bottom panel) are the results of minimizing, 
respectively, $e_1$ and $e_2$ which for $T=2~\frac{2\pi}{\delta\omega}$ both are 
of the order of $10^{-12}$. The corresponding Gaussian pulse is also shown 
in both panels. $g_1$ is guaranteed to decrease the error of NOT-gate two orders 
of magnitude with respect to the Gaussian pulse while $g_2$ would reduce the error 
up to ten orders of magnitude.

\begin{figure}[top]
\begin{center}
\begin{tabular}{cc}
\includegraphics[width=1.0\linewidth]{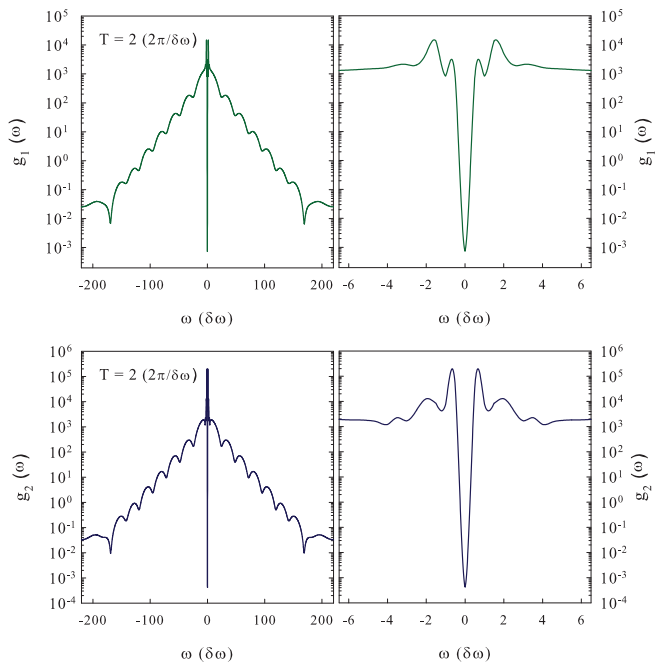}
\end{tabular}
\end{center}
\caption{(Color online) Fourier transform $g(\omega)$ of two optimized pulses shown in
Fig.~\ref{fig:pulses} (dashed lines). $g_1(\omega)$ minimizes $e_1$ and 
$g_2(\omega)$ minimizes $e_2$. $\delta\omega$ is chosen to be ten percent 
of $\omega_{01}$ and $\omega_{01}/2\pi$ is approximately $5~\text{GHz}$.
\label{fig:fourier}}
\end{figure}

\begin{figure}[top]
\begin{center}
\includegraphics[width=1.0\linewidth]{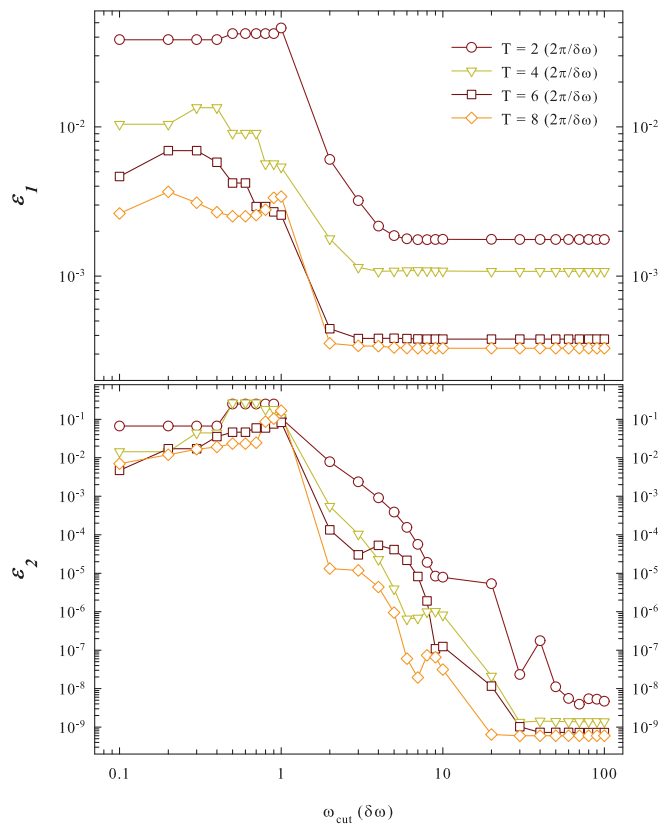}
\end{center}
\caption{(Color on line) Error $\cal E$ for a NOT-gate with optimized pulses 
obtained by minimizing $e_1$ (top panel) and $e_2$ (bottom panel) as a function 
of the cutoff frequency $\omega_{cut}$. Integer values of $T\omega_{cut}/2\pi$ 
correspond to the number of Fourier components included. $\delta\omega/2\pi$ is 
approximately $500$ MHz.
\label{fig:cutoff}}
\end{figure}

Fig.~\ref{fig:fourier} shows the Fourier transform of the two optimized pulses shown in
Fig.~\ref{fig:pulses}. To filter out the high-frequency components of the 
optimized pulses we set a cutoff frequency $\omega_{cut}$ for Fourier components 
and apply the truncated pulses again and obtain the error. 
Fig.~\ref{fig:cutoff} shows the error for a NOT-gate for pulses with different 
duration times as functions of $\omega_{cut}$. $\omega_{01}/2\pi$ is approximately $5$ GHz
and $\delta\omega$ is typically $10\%$ of $\omega_{01}$. 
In our calculation $\delta\omega=0.1~\omega_{01}$ which means that
$\delta\omega/2\pi\approx~500$ MHz.

In the case of ${\cal{E}}_1$, top panel of figure~\ref{fig:cutoff} makes clear 
that all important harmonics have frequencies smaller than $5\delta\omega$.
Note that the number of harmonics included within the cutoff is equal to 
$T\omega_{cut}/2\pi$ so that, for $T=N(2\pi/\delta\omega)$, such number 
is equal to N times the ratio $\omega_{cut}/\delta\omega$. As a result it 
seems that, for all values of $T$ considered, about 20 harmonics should be 
sufficient to reach the smallest value of ${\cal{E}}_1$. ${\cal{E}}_2$, though, 
seems to be more sensitive to high-frequency components but still about four 
orders of magnitude smaller than ${\cal{E}}_1$ under the cutoff 
$\omega_{cut}=10\delta\omega$.

\subsection{Effect of off-resonance terms}
\label{off}

\begin{figure}[t]
\begin{center}
\includegraphics[width=1.0\linewidth]{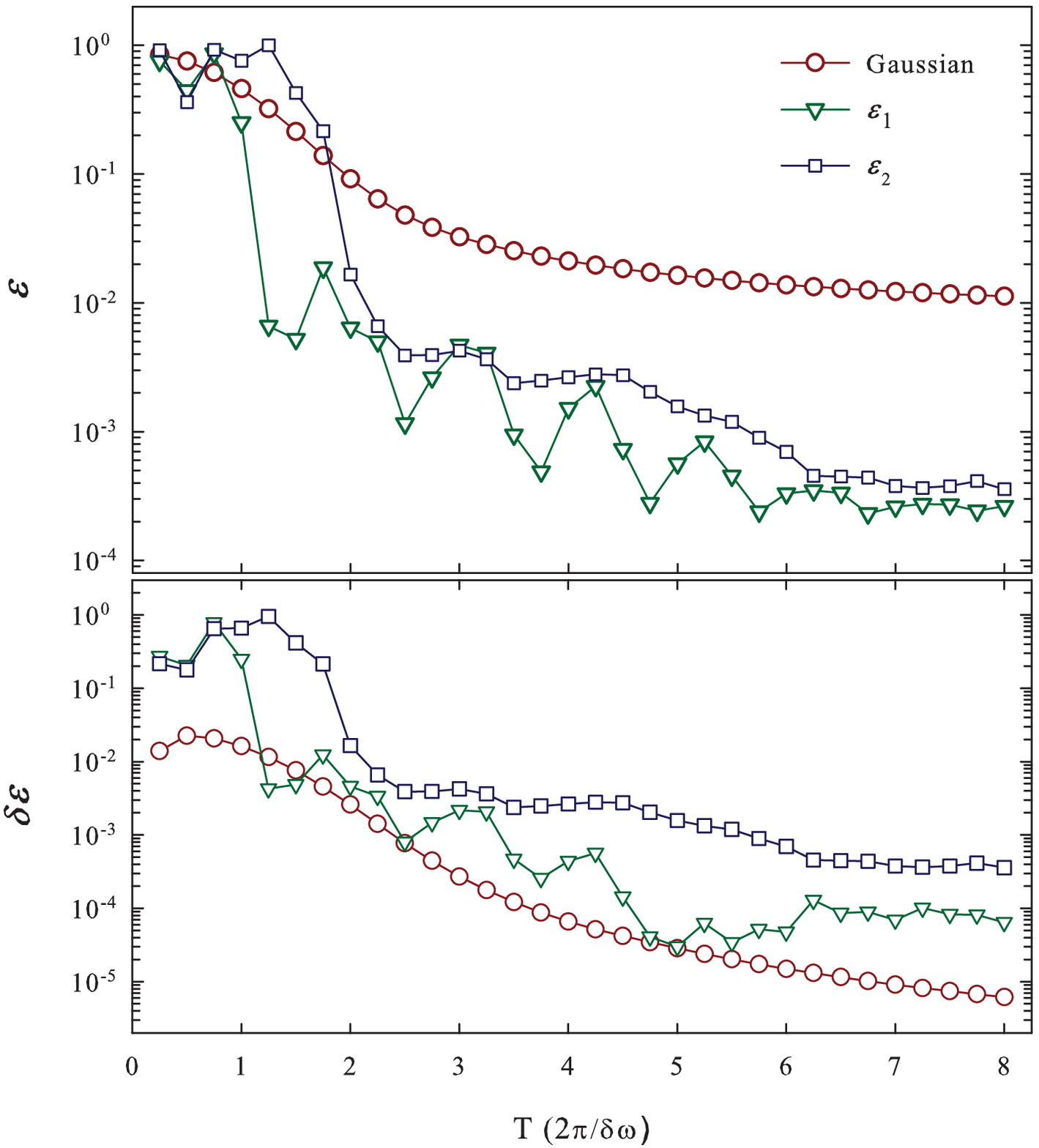}
\end{center}
\caption{(Color on line) Top panel: the error $\cal E$ for a NOT-gate
made by applying microwave pulses with Gaussian modulation (circles) and 
optimized modulation (triangles and squares) when off-resonance terms are kept.
Note that optimized pulses are obtained by excluding off-resonance elements.
Bottom panel: the absolute value of error difference $\delta{\cal{E}}$ 
obtained by subtracting the curves in the top panel from those in 
Fig.~\ref{fig:optimized}. $T$ is the total time width of the pulses.
\label{fig:off-res}}
\end{figure}

As we mentioned before, we have assumed that off-resonance elements of the 
Hamiltonian (\ref{eq:lab-to-rot}) in the rotating frame are negligible
and we have used Hamiltonian (\ref{eq:rotating2}) for calculating the evolution.
In this section we check this assumption by addressing the effect of off-resonance 
elements by evolving the complete Hamiltonian (\ref{eq:lab-to-rot}) 
using the optimized pulses obtained using Hamiltonian (\ref{eq:rotating2}).
Top panel of Fig.~\ref{fig:off-res} shows the error for a NOT-gate 
operation implemented by Gaussian (circles) and optimized pulses from minimizing 
$e_1$ (triangles) and $e_2$ (squares). For $T>2~\frac{2\pi}{\delta\omega}$ the optimized 
pulses yield a much higher error, with respect to the case when off-resonance terms
are neglected, still showing an improvement of two orders of magnitude if compared 
to Gaussian pulses. Bottom panel of Fig.~\ref{fig:off-res} shows the absolute value 
of the error difference $\delta {\cal E}$ obtained by subtracting the error without 
off-resonance term from the error with off-resonance terms.
These figures make clear that while for Gaussian pulses off-resonance terms can be neglected, 
for optimized pulses, specially those obtained from minimizing $e_2$, they are very 
important. Note that, contrary to the ideal case where ${\cal{E}}_2$ was about eight 
orders of magnitude smaller than ${\cal{E}}_1$, under the effect of off-resonance terms, 
${\cal{E}}_2$ seems to be larger than ${\cal{E}}_1$ specially for very short pulses with 
$T<2~\frac{2\pi}{\delta\omega}$. This means that the assumption of ignoring these terms 
is more accurate when $e_1$ is minimized. The simpler shape of the optimized pulses obtained 
from minimization of $e_1$ could be a reason for that.

\subsection{Effect of capacitive interaction}
\label{int}

So far we have considered a single qubit with three energy levels and obtained
the modulation of the microwave pulses in order to optimize the NOT-gate
operation for the two lowest energy states $|0\rangle$ and $|1\rangle$.
It is now interesting to consider the setup \cite{SteffenSCI06} containing two
qubits interacting via a capacitor. The question that we want to address is 
what happens if these optimized pulses are applied on the first qubit while 
the interaction with the second qubit is present.

\begin{figure}[top]
\begin{center}
\includegraphics[width=1.0\linewidth]{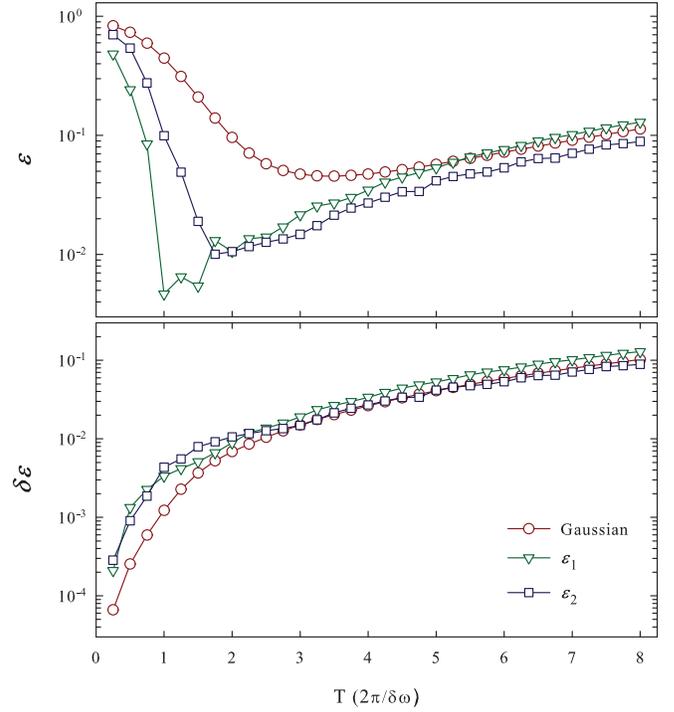}
\end{center}
\caption{(Color on line) Error ${\cal{E}}$ for a NOT-gate on the first qubit
implemented by applying Gaussian (circles) and optimized (triangles and squares) 
pulses on the first qubit of a two-qubit system in presence of capacitive interaction. 
Gaussian pulses have amplitude $a=1.25$ and cutoff in time $\alpha=3$ and are used 
as initial guess in the optimization procedure. Optimized pulses are obtained from 
the single-qubit setup. The strength of the interaction is $C_x/C_{\Sigma}=2.3\times 10^{-3}$.
Bottom panel: the absolute value of error difference $\delta{\cal{E}}$ obtained by 
subtracting the curves in the top panel from those in Fig.~\ref{fig:optimized}.
\label{fig:interaction}}
\end{figure}

The interaction Hamiltonian of a circuit with two identical phase qubits has 
the following form:

\begin{eqnarray}
\label{eq:2bitHint}
H_{int} = - \frac{E_C^2}{E_{C_x}}\left[ 
\left(\frac{\partial^2}{\partial\delta_1^2}+\frac{\partial^2}{\partial\delta_2^2}\right)
+ 2\left(i\frac{\partial}{\partial\delta_1}\otimes i\frac{\partial}{\partial\delta_2}\right)\right]
\end{eqnarray}
where $\delta_1$ and $\delta_2$ are Josephson phases across the junction 
$1$ and $2$ and $C_x$ is the capacitance of the interaction capacitor.
Note that the term with second derivative in Eq.~(\ref{eq:2bitHint}) can 
be included in the Hamiltonians of the uncoupled qubits (\ref{eq:washboard}) 
by replacing the charging energy $E_C$ with an effective one 
$E_{C_{eff}}=(2e)^2/(2C_{eff})$, where $C_{eff}\equiv C^2/C_{\Sigma}$ and 
$C_{\Sigma}=C+C_x$. The Hamiltonian can again be written in the basis of the 
eigenstates of the uncoupled qubits, and the strength of the
interaction Hamiltonian reduces to 
 $(C_x/C_{\Sigma})(\hbar\omega_{01})$.
We  move to the rotating frame described by the unitary operator

\begin{eqnarray}
\label{eq:2bit-unitary}
V &=&
\left(\begin{matrix}
1&0&0\\
0&e^{i \omega t}&0\\
0&0&e^{2i \omega t}
\end{matrix}\right )
\otimes
\left(\begin{matrix}
1&0&0\\
0&e^{i \omega t}&0\\
0&0&e^{2i \omega t}
\end{matrix}\right )
\end{eqnarray}
and neglect the off-resonance elements of the resulting Hamiltonian. By applying 
microwave pulse on the first qubit, our aim is to perform a NOT-gate operation on 
such qubit (namely, $\sigma_{x1}\otimes\mathds{1}_2$). Since $C_x$ is typically 
of few fF and $C$ is of the order of pF~\cite{SteffenPRL06, SteffenSCI06}
we find $C_x/C_{\Sigma}\simeq 2.3\times 10^{-3}$ which leads to an interaction 
strength $(C_x/C_{\Sigma})\omega_{01}\approx 10$MHz. Figure \ref{fig:interaction} shows 
the error as function of time width of the pulse $T$ for both Gaussian (circles) and 
optimized pulses ( triangles and squares). Although the optimized pulses are obtained for a 
single qubit system, they still result in smaller error at least for short pulses. 
These results show the importance of the presence of the capacitive interaction 
even though the strength of the interaction is small. As it is clear from 
Fig.~\ref{fig:interaction} for pulses longer than, approximately, $8$ ns 
($T=4~\frac{2\pi}{\delta\omega}$) the interaction becomes more effective and 
the error for Gaussian and optimized pulses are very close. Moreover, longer pulses 
lead to higher value of error contrary to what happens in the case of a single qubit.
In this case also for very short pulses ($T<1.75~\frac{2\pi}{\delta\omega}$) 
${\cal{E}}_1$ is smaller than ${\cal{E}}_2$ and for $T\geq 2~\frac{2\pi}{\delta\omega}$ 
they are of the same order, although in the ideal case ${\cal{E}}_1$ was eight orders 
of magnitude larger than ${\cal{E}}_2$.
 
Bottom panel of Fig.~\ref{fig:interaction} shows the absolute value of error 
difference $\delta{\cal{E}}$ which is obtained by subtracting the error of ideal 
case from the error in presence of interaction. $\delta{\cal{E}}$ is almost the 
same in all three cases.

\section{Leakage}
\label{five}

As explained in section \ref{model}, the two lowest energy levels of a 
current-biased Josephson junction can be used as $|0\rangle$ and $|1\rangle$ 
states of a phase qubit. Although it would be desirable to have only 
two levels inside the potential well of the Fig.~\ref{fig:junction},  this 
is not the case in experimental setups. So far we have included the leakage 
by considering only an additional third level and showed that it is possible 
to optimize the pulses in order to gain high fidelity for a NOT-gate for a single qubit.
In typical experiments the number of energy levels inside the well varies 
between three and five. In order to have a more complete understanding of the leakage, 
in this section we show some results obtained for a five-level system.
Since adding more levels to the system decreases the inhomogeneity of the level-spacing
we choose $\delta\omega=0.05\omega_{01}$.

\begin{figure}[top]
\begin{center}
\includegraphics[width=1.0\linewidth]{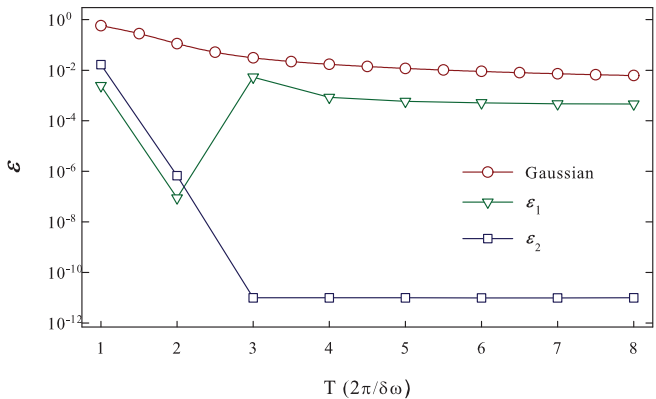}
\end{center}
\caption{(Color on line) The error $\cal E$ of NOT-gate made by applying 
Gaussian (circles) and optimized (triangles and squares) pulses as function of duration 
time of the pulse $T$. Gaussian pulses have amplitude $a=1.25$ and cutoff in time 
$\alpha=3$ and are used as initial guess in optimization. Optimized pulses are obtained 
after at most $15000$ iterations. Physical system contains five energy states and 
$\delta\omega$ is assumed to be $0.05~\omega_{01}$.
\label{fig:5-optimized}}
\end{figure}

\begin{figure}[top]
\begin{center}
\includegraphics[width=1.0\linewidth]{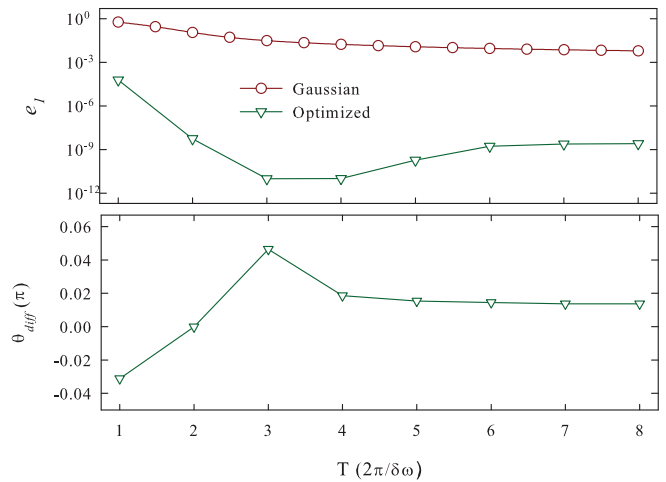}
\end{center}
\caption{(Color on line) Top panel: the averaged value of $e_1$ for 
$|0\rangle\leftrightarrow|1\rangle$ transitions after applying the Gaussian 
pulses (circles) and optimized pulses (triangles) obtained from minimizing $e_1$. 
Vertical axis is in logarithmic scale. Bottom panel: the phase difference between 
final $|0\rangle$ and $|1\rangle$ states in units of $\pi$.
\label{fig:5-phasediff}}
\end{figure}

Figure~\ref{fig:5-optimized} shows the error $\cal E$ for a NOT-gate implemented 
by optimized and Gaussian pulses, which are used as initial guess, 
for different duration times $T$. Similar to the case of three-level system, with 
same number of iterations, minimization of $e_1$ leads to better results for short 
pulses while for longer duration times of pulses minimizing $e_2$ results in error 
of NOT-gate ${\cal{E}}_2\approx 10^{-12}$. In the case of minimizing $e_1$, at least one 
order of magnitude improvement is achieved for long pulses, although the improvement 
obtained for pulses with shorter time width are the best. By looking at the average 
value of $e_1$ for transitions between the states $|0\rangle$ and $|1\rangle$ 
(top panel of Fig.~\ref{fig:5-phasediff}) and the final phase difference between them 
(bottom panel of Fig.~\ref{fig:5-phasediff}), one realizes that, as it was observed 
in three-level system, considerable amount of ${\cal{E}}_1$ is due to the final phase 
difference $\theta_{diff}$. For instance the pulse with $T=2~\frac{2\pi}{\delta\omega}$ 
results in a phase difference approximately equal to zero and therefore ${\cal{E}}_1$ 
for this pulse is of the order of $10^{-8}$.
A proper phase shift applied after the NOT-gate operation will compensate the 
phase difference between the final $|0\rangle$ and $|1\rangle$ 
states and consequently attaining a very high fidelity.

\begin{figure}[top]
\begin{center}
\includegraphics[width=1.0\linewidth]{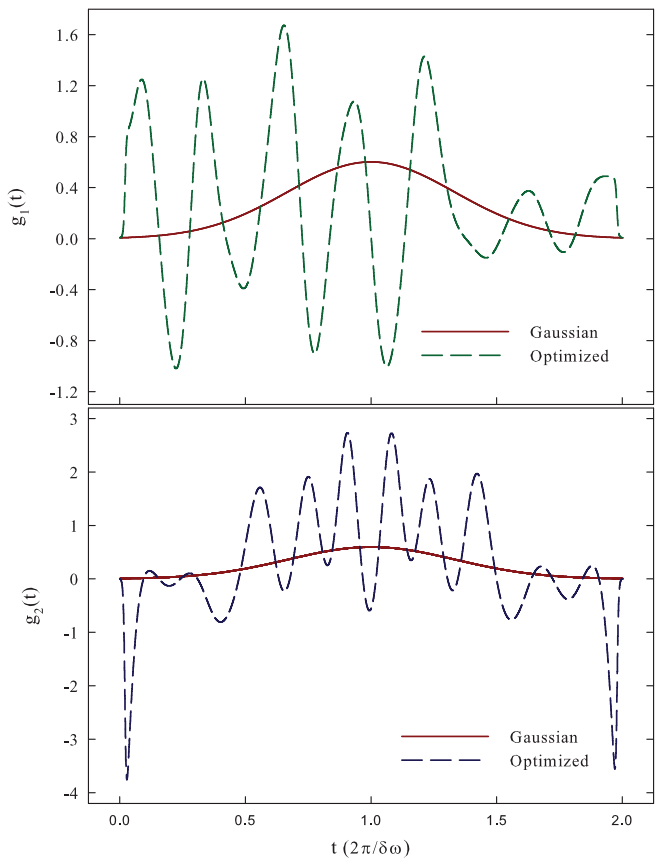}
\end{center}
\caption{(Color on line) Examples of final optimized modulation of pulses 
(dashed lines), obtained from minimizing $e_1$ (top panel) and $e_2$ 
(bottom panel), and the corresponding pulse with Gaussian modulation 
(solid lines) used as initial guess in optimization process with duration 
time $T=2~\frac{2\pi}{\delta\omega}$ in a system with five energy states. 
\label{fig:5-pulses}}
\end{figure}

Two examples of pulses with $T=2~(2\pi/\delta\omega)$ are shown in figure 
\ref{fig:5-pulses}. $g_1$ is obtained from minimizing $e_1$ and gives rise to 
${\cal{E}}_1\approx 10^{-8}$ while $g_2$ is supposed to minimize $e_2$ with 
${\cal{E}}_2\approx 10^{-7}$. It seems that, compared to three-level system, 
higher frequencies and amplitudes are needed to reach high fidelity of NOT-gate.
In three-level system the iterative optimization algorithm is applied 
at most $5000$ times to reach such fidelities while with five levels $15000$ 
iterations were needed to obtain the results shown in figures \ref{fig:5-optimized} 
and \ref{fig:5-phasediff}. The leakage out of the qubit manifold would be the reason 
for this.

\section{Conclusions}
\label{conc}
In this paper we have shown that it is possible to optimize 
single-qubit gates for Josephson phase qubits by employing quantum optimal 
control theory.
We have considered the realistic situation in which, in addition to the two 
computational basis states $|0\rangle$ and $|1\rangle$, higher energy states 
are present, which may lead to leakage.
Typically microwave pulses with Gaussian modulation are used to induce 
transition between states $|0\rangle$ and $|1\rangle$, yielding a quite 
high fidelity for long pulse durations.
For the sake of definiteness, here we have focused on the NOT-gate 
single-qubit operation and searched for modulations of microwave pulses 
which optimize such operation, especially for short-duration pulses.
The numerical results obtained for a three-level system, and neglecting 
off-resonance terms, demonstrate up to ten orders of magnitude improvement 
in fidelity of a NOT-gate operation with respect to those obtained through 
Gaussian modulations. To test the effect of possible imperfections in the 
pulses shape, we have studied the behavior of the fidelity as a function 
of the bandwidth of the pulse generator and showed that frequencies not 
bigger $2$ GHz are needed to gain up to four orders of magnitude improvement.
Moreover, we have shown that the off-resonance elements of the Hamiltonian, 
which are usually neglected, can be important for optimized pulses, especially 
for short pulse duration times, due to the very high fidelity reached. 
We have also addressed the effect 
of the presence of a capacitively-coupled second qubit and showed that, 
even though the optimized pulses are obtained for a single qubit, they 
still lead to a high fidelity for a NOT-gate (up to two orders of 
magnitude improvement) especially for very short pulses.
Finally, we were able to obtain optimized pulses 
for a system with 5 energy levels in which the leakage outside 
qubit manifold is more severe.

In conclusion, the two-interacting-qubit system deserves for sure further attention. 
On the one hand, in order to improve the fidelity of a single-qubit operation, 
in the presence of capacitive coupling, it seems that a way to switch the 
interaction on and off should be found even for optimized pulses of an isolate qubit.
On the other hand, obtaining optimized pulses while including the interaction, 
would be a potential theoretical work to be done.

\acknowledgments
We would like to acknowledge fruitful discussions with F. W. J. Hekking.
We acknowledge support by EC-FET/QIPC (EUROSQIP) and by
Centro di Ricerca Matematica ``Ennio De Giorgi'' of Scuola Normale
Superiore. S. M. acknowledges support by EU-project SCALA.

\end {document}